\input psfig
\documentstyle[12pt]{article}
\newskip\humongous \humongous=0pt plus 1000pt minus 1000pt
\def\caja{\mathsurround=0pt}

\def\eqalign#1{\,\vcenter{\openup1\jot \caja
        \ialign{\strut \hfil$\displaystyle{##}$&$
        \displaystyle{{}##}$\hfil\crcr#1\crcr}}\,}
\newif\ifdtup

\def\be{\begin{equation}}
\def\ee{\end{equation}}
\def\ba{\begin{eqnarray}}
\def\ea{\end{eqnarray}}

\begin{document}
\renewcommand{\theequation}{\thesection.\arabic{equation}}
\newcommand{\beq}{\begin{equation}}
\newcommand{\eeq}[1]{\label{#1}\end{equation}}
\newcommand{\ber}{\begin{eqnarray}}
\newcommand{\eer}[1]{\label{#1}\end{eqnarray}}
\begin{titlepage}
\begin{center}

\hfill CPTH-S4540696 \\
\hfill Crete-96-11   \\
\hfill hep-ph/9606348\\

\vskip .5in

{\large \bf ON SPHERICALLY-SYMMETRIC
 SOLUTIONS IN THE TWO-HIGGS STANDARD MODEL} 
\vskip .5in

{\bf C. Bachas$ \ ^{\spadesuit}$, P. Tinyakov$ \ ^{\diamondsuit}$ 
and T.N. Tomaras$ \ ^{\clubsuit}$} 
\vskip .1in

{
$\ ^{\spadesuit}$
 Centre de Physique Th\'eorique, Ecole Polytechnique,
 91128 Palaiseau, FRANCE \\
 email: {\it bachas@orphee.polytechnique.fr}\\
\   \\
$\ ^{\diamondsuit}$
  Institute for Nuclear Research, 
60th October Anniversary prospect 7a,
 117 312 Moscow, RUSSIA\\
email: {\it tinyakov@ms2.inr.ac.ru} \\
 \    \\
$\ ^{\clubsuit}$
Dept. of Physics and Institute of Plasma Physics, 
University of Crete, \\
and Research Center of Crete,\\
P.O. Box 2208, 710 03 Heraklion, GREECE\\
email: {\it tomaras@physics.uch.gr}
}

\vskip .15in

\end{center}

\vskip .4in

\begin{center} {\bf ABSTRACT }
\end{center}

\vskip .15in

\begin{quotation}

\noindent

We report the results of a numerical search for non-topological
solitons in the two-Higgs standard model, characterized by the
non-trivial winding, $\pi_3(S^3)$, of the relative phase of the two
doublets.  In a region of (weak-coupling) parameter space we identify
a branch of winding solutions, which are lower in energy than the
embedded standard sphaleron and deformed (bi)sphalerons.  Contrary,
however, to what happens in 2d toy models, these solutions remain
unstable even for very large Higgs masses.

\end{quotation}
\end{titlepage}
\vfill
\eject
\def\baselinestretch{1.2}
\baselineskip 16 pt
\noindent
\setcounter{equation}{0}

Two of us have analyzed recently non-topological winding solitons in
renormalizable gauge models with two Higgs scalars.  Such solitons
arise in two and three space-time dimensions \cite{BT1}, and can be
embedded as extended membrane defects in the weakly-coupled two-Higgs
standard model (2HSM) \cite{BT3}.  It was furthermore argued by
analogy \cite{BT1} that the 2HSM may have string-like or particle-like
excitations characterized by a non-trivial mapping of the spatial
three sphere onto the relative phase of the two doublets. Depending on
the details of the potential, the relevant quantity could be either
the Hopf invariant, $\pi_3(S^2)$, or the winding number, $\pi_3(S^3)$.
Here we report the results of a numerical search for solitons of the
latter type. The search revealed indeed a branch of winding solutions,
but unlike their 2d analogs these stay classically unstable for
arbitrarily large Higgs masses. These new solutions could nevertheless
be of interest for the study of electroweak baryogenesis in the 2HSM
\cite{Turok}.  Sphaleron solutions in the 2HSM have been discussed
previously in ref. \cite{Peccei} and in a different context in
ref. \cite{Bri}.

    There exists a close analogy between sphaleron solutions in the
minimal standard model \cite{spha,defspha,Yaffe}, and those in the
abelian-Higgs \cite{2dab} or global double-well \cite{2d} models on
the circle.  Similarly, the structure of classical solutions in the
2HSM is strongly reminiscent of that in the abelian two-Higgs, or in
the global mexican-hat model on the circle. This latter model is
defined by the action 
$$ S = \int dt \int_0^{2\pi L} dx
\Biggl[ {1\over 2} \partial_\mu\Phi^* \partial^\mu\Phi 
-{\lambda \over 4}(\Phi^*\Phi -  v^2)^2 \Biggr] \ , \eqno(1)
$$
whose single classically-relevant parameter is $mL=
\sqrt{2\lambda} vL$. We henceforth set $L=1$.
Static solutions correspond to motions of a particle in the inverted
mexican-hat potential with period $2\pi $. One solution, which exists
for all values of $m$ and can be thought of as the basic sphaleron
\footnote{Though it does  not in this case
represent a minimum-height passage between two discrete degenerate
vacua.}, corresponds to the particle sitting at the bottom of the
inverted hat.  It has energy $\pi m^4 / 8\lambda$, starts out with two
negative modes, and acquires two extra ones every time $m/\sqrt{2}$
crosses a positive integer $n$.  Two extra branches of solutions
bifurcate at these points: the $S_n$ branch describes an n-fold
oscillation about the bottom of the inverted hat, with zero angular
momentum.  These solutions, also present in the double-well model, are
everywhere unstable and are the analogs of the deformed sphalerons of
the standard model \cite{2d,defspha}. The $W_n$ branches on the other
hand describe an $n$-fold rotation around the tip of the hat, and have
no analog in the one-Higgs model.  The solutions have a constant
scalar magnitude $\Phi^*\Phi= (m^2-2n^2) / {2\lambda}$, an energy
equal to ${\pi n^2} (m^2-n^2) / {2\lambda}$, and become classically
stable beyond the critical values $m^2 = 6n^2-1$ \cite{BT1}.  New
branches, ${\tilde W}_n$, emerge at these points; they have slightly
higher energy and correspond to winding motions with an oscillating
scalar magnitude.  Further bifurcations occur along the $W_{n>1}$
branches, and possibly also along $S_n$ and ${\tilde W}_n$, but an
exhaustive analysis is beyond the scope of this letter.  Figure 1 is
an illustration of the topological tree of solutions. The main lesson
to retain
\vskip 0.3cm
\centerline{\hbox{
 \psfig{file=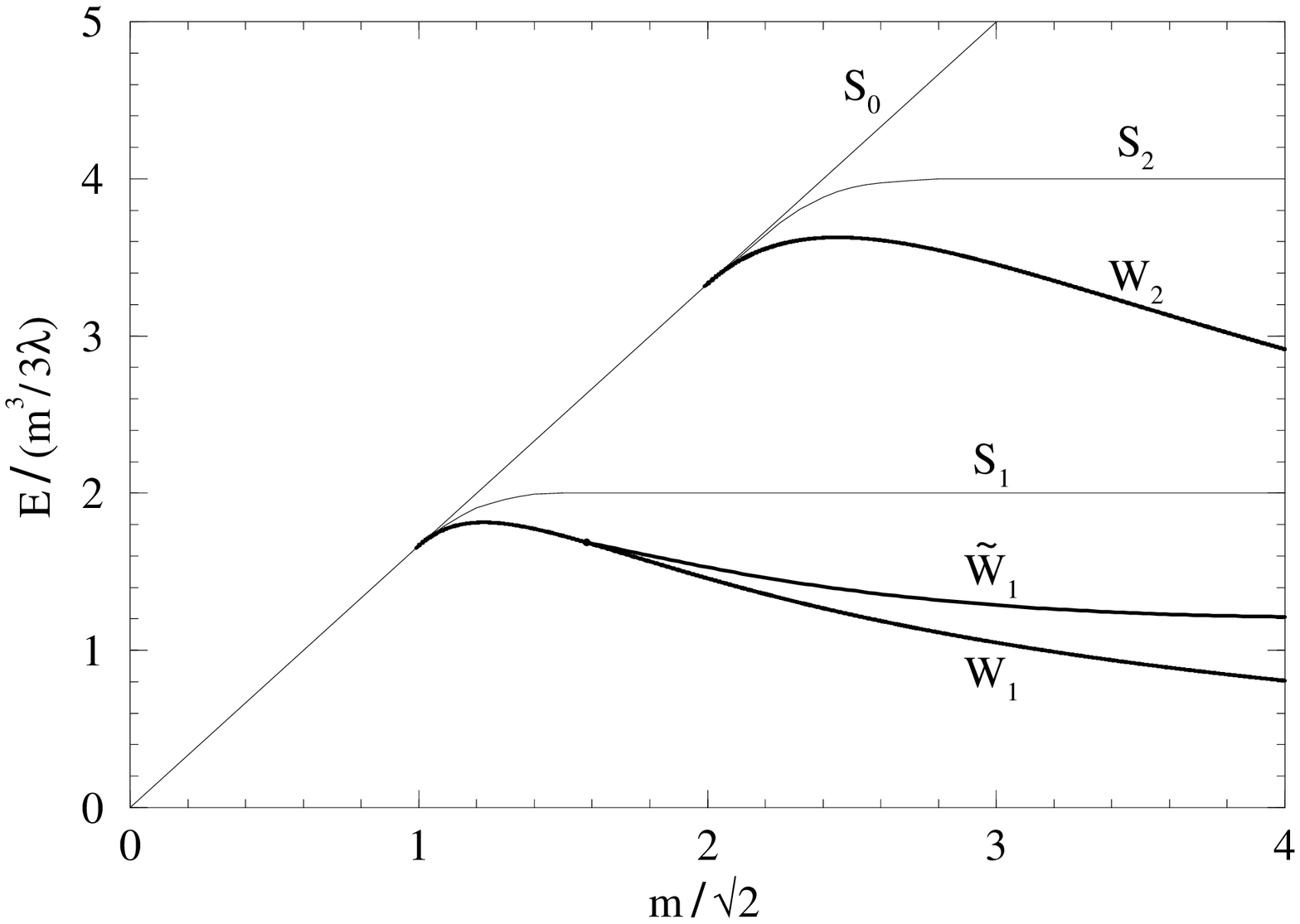,height=8cm,width=12.5cm,angle=270}
 \hskip 1.2cm
}}
\vskip 0.4cm
\font\ninerm=cmr9

{\bf Figure 1}.{ \ninerm Part of the
tree of classical solutions for the model (1).
The vertical axis gives the energy, measured  in units of }
$\scriptstyle m^3 / {3 \lambda}$
{\ninerm  so as  to 
simplify comparison with ref. \cite{2d}. The}
$\scriptstyle {\tilde W}_1$ 
{\ninerm  branch
is drawn out of scale:  its energy differs by only } 
$\scriptstyle \sim 1\%$ 
{\ninerm  from
the energy of} 
$\scriptstyle  W_1$.
\eject

\noindent
from this toy model is that for $m>\sqrt{2}$ the winding solution on
the $W_1$ branch is the one with lowest energy, and that furthermore
for $m>\sqrt{5}$ it becomes a classically-stable soliton.

Using this toy model as a guide, we turn now to the 2HSM.  We will
work with the simplifying assumptions of vanishing Weinberg angle and
of a potential consisting of two independent mexican hats.  We believe
that these restrictions are optimal, in the sense of favouring the
stability of the sought-for winding solutions. We also restricted our
search to spherically-symmetric configurations.  The action is
$$\eqalign{
  S={1\over {g^2}}\int  
d^4x \Biggl[ -{1\over 2} Tr(W_{\mu\nu}&W^{\mu\nu}) 
  +\sum_{a=1,2} (D_{\mu} H_a)^{\dagger}(D^{\mu} H_a)\cr &
  -\sum_{a=1,2} {{\lambda_a}\over {g^2}} 
  (H^{\dagger}_a H_a- g^2 v^2_a)^2 \Biggr]\cr} \eqno(2)
$$
with $ D_\mu H_a=(\partial_\mu + W_\mu) H_a$, for $a=1,2 $, and
$W_{\mu\nu}= [D_\mu, D_\nu]$. We fix the length scale by setting
$m_W=g \sqrt{(v_1^2+v_2^2)/2}=1 \ , $ and scale out of the action the
semiclassical parameter $\alpha_W = g^2/4\pi$.  There remain three
classically-relevant parameters: the two radial Higgs masses $m_a =
2\sqrt{\lambda_a} v_a$, and ${\rm tan}\beta = {v_2}/{v_1}$.

  The general  spherically-symmetric static ansatz,
following the conventions of Yaffe \cite{Yaffe},
reads
$$\eqalign{
 & W_0=w_0\tau_in_i/2i\; ,
  \cr 
  W_i={1\over 2i}\ \Bigl[ ({\rm a}-1)&\epsilon_{ijk}\tau_jn_k/r + 
      {\rm b} (\delta_{ij}-n_in_j)\tau_j/r
 + w_1n_in_j\tau_j \Bigr]
  \cr
  &
  H_a=(\mu_a+i\nu_a n_i\tau_i)\xi 
  \cr} \eqno(3)
$$

where $n_i$ is the unit vector in the direction of $\bf r$, $\tau_i$
are the three Pauli matrices, $\xi$ is a constant unit doublet and
$w_0, w_1, {\rm a}, {\rm b}, \mu_a, \nu_a$ are functions of $r$ to be
determined by the stationarity of the energy.  It is convenient to
choose the gauge $w_0=0$, and then solve Gauss' constraint to obtain
$w_1 = 0$ \cite{Yaffe}.  Defining the complex fields
$\phi_a=\mu_a+i\nu_a \equiv f_a \,{\rm exp} \,(i\theta_a)$ ($a=1,2$)
and $\chi = {\rm a} + i {\rm b} = \vert\chi \vert \,{\rm exp}
\,(i\psi)$, one can put the energy functional in the form:
$$\eqalign{
  E = & {1 \over \alpha_W} 
    \int_0^\infty  dr \Biggl\{
  \vert \chi^\prime \vert^2  
  + 2r^2 {\rm cos}^2\beta
  (\vert \phi_1^\prime \vert^2+\vert \phi_2^\prime
\vert^2 )
+ {1\over 2r^2}(\vert\chi\vert^2-1)^2
  \cr
  &
  + {\rm cos}^2\beta (\vert\chi\vert^2+1)(\vert\phi_1\vert^2+
\vert\phi_2\vert^2) 
 - 2 {\rm cos}^2\beta\  Re[\chi^* (\phi_1^2+\phi_2^2)]
  \cr
  &+ {1\over 2} m_1^2 r^2 {\rm cos}^2\beta 
 (\vert\phi_1\vert^2-1)^2
  + {1\over 2} m_2^2  r^2
{{\rm cos}^2\beta\over {\rm tan}^2\beta} 
(\vert\phi_2\vert^2-{\rm tan}^2\beta)^2 \Biggr\},
  \cr} \eqno(4)
$$
with primes standing for derivatives with respect to $r$.  The above
expression is invariant under a simultaneous rotation of $\phi_1$,
$\phi_2 $ and $\chi$ by a phase $\omega, \omega $ and $ 2 \omega$
respectively, as well as under independent changes of sign of $\phi_1$
and $\phi_2$. These symmetries together with the requirement of finite
energy allows us to fix the boundary conditions at $r=\infty$ in the
form:
$$
\chi\rightarrow 1 \ , \  \phi_1 \rightarrow 1 \ \ {\rm and} \ \
\phi_2 \rightarrow {\rm tan}\beta \ . \eqno(5)  
$$
At $r=0$ on the other hand, finiteness of the energy, of the gauge
current and of the field strength imply \cite{Yaffe}:
$$
\vert\chi\vert \rightarrow 1 \ , \   
\chi^*\phi_a^2 \rightarrow \vert\phi_a\vert^2 \ \
{\rm and} \ \ \chi^\prime \rightarrow 0 \ .  \eqno(6) 
$$
Notice in particular that at the origin either $f_a = 0$, or else
$2 \theta_a -\psi = 0 \;({\rm mod} \;2 \pi)$. 
For configurations with non-vanishing 
Higgs magnitudes we may conclude that
$$
\theta_1 -  \theta_2 \Biggl\vert_{r=0}\ =  {\rm N} \pi \eqno(7)
$$
The integer N is the gauge-invariant winding number characterizing the
mapping from the spatial $S^3$ onto the SU(2) manifold of the relative
phase of the two doublets \cite{BT1}.

We used the relaxation method \cite{Numerics} to integrate the field
equations numerically. Starting with some initial guess $\{\hat\Phi\}$
for the solution, one linearizes the static field equations, solves
them, and adds the result to $\{\hat\Phi\}$ thus obtaining an improved
guess for the next iteration.  The radial coordinate $r$ was replaced
by n+1=201 points $r_i$, with $r_0=0$ and $r_n=R=20$.  This was
sufficient to achieve an accuracy of order $10^{-3}$ to $10^{-4}$ in
the energy of the solutions.  Following ref. \cite{Yaffe} we
distributed the lattice points according to the formula $r_k=(1-{\rm
e}^{k s/n}) / ({\rm e}^{k s/n} - \mu)$, with $s={\rm ln} [(1+\mu
R)/(1+R)]$ and $\mu={\rm max} (m_1, m_2)$.  The linearized equations
at each iteration step take the form $E^{(2)}_{IJ} \delta\Phi^J =
E^{(1)}_I$ where $\delta\Phi$ is a vector of dimension $n\times(\#
\mbox{fields})=1200$, normalized so that the kinetic energy reads
${1\over 2} \sum\dot {\delta\Phi^i} \, \dot {\delta\Phi^i} $, and
$E^{(1)}$, $E^{(2)}$ are respectively the first and second variations
of the static energy, eq. (4).  The matrix $E^{(2)}$ has a special
block-three-diagonal form, each block corresponding to the six fields
at the same lattice point. We have exploited this special form to
invert the matrix by a forward-elimination and back-substitution
procedure. This same procedure also allowed us to calculate the
spectrum of negative fluctuation modes around the solutions.

Figure 2 summarizes schematically the tree of solutions that we have
found in the particular case of equal Higgs masses, $m_1=m_2\equiv m$.
This further restriction of parameter space has the following
convenient feature: any solution $\{ {\hat H}, {\hat W}_\mu \}$ , of
the one-doublet model with Higgs mass $m$ and $m_W=1$, yields a
solution $\{ H_1 = {\hat H} {\rm cos}\beta , \ H_2 = {\hat H} {\rm
sin}\beta , \ W_\mu = {\hat W}_\mu \}$ of the 2HSM with the same total
energy though a different fluctuation spectrum. The branches $S_0$ and
$S_1$ in particular correspond to the standard sphaleron \cite{spha}
and deformed (bi)sphalerons \cite{defspha,Yaffe} of the minimal model,
and have been used to check the accuracy of our numerical routines.
All other branches shown in figure 2 have no analog in the one-doublet
model.

The standard sphaleron (with $\chi, \phi_a$ real) is the only solution
we have found for $m/m_W \le 5.5$. It has a single mode of instability
in this range, but develops a second one where the new branches of
winding solutions, ($W_1$) and its conjugate, bifurcate. Extra
negative modes appear at each subsequent bifurcation: at $m/m_W \simeq
12$ (first deformed sphalerons), at $m/m_W \simeq 50$ (doubly-winding
solutions), at $m/m_W \simeq 138 $ (second deformed sphalerons) and so
on down the line. This sequence of bifurcations is strongly
reminiscent of the 2d toy model, except that the $W_n$ and $S_n$
branches do not emerge in this case simultaneously.  Notice that the
topological invariant, $\sum_{\rm solutions} (-)^{n_s}$,

\eject

\vspace{0.6cm}
\centerline{\hbox{
 \psfig{file=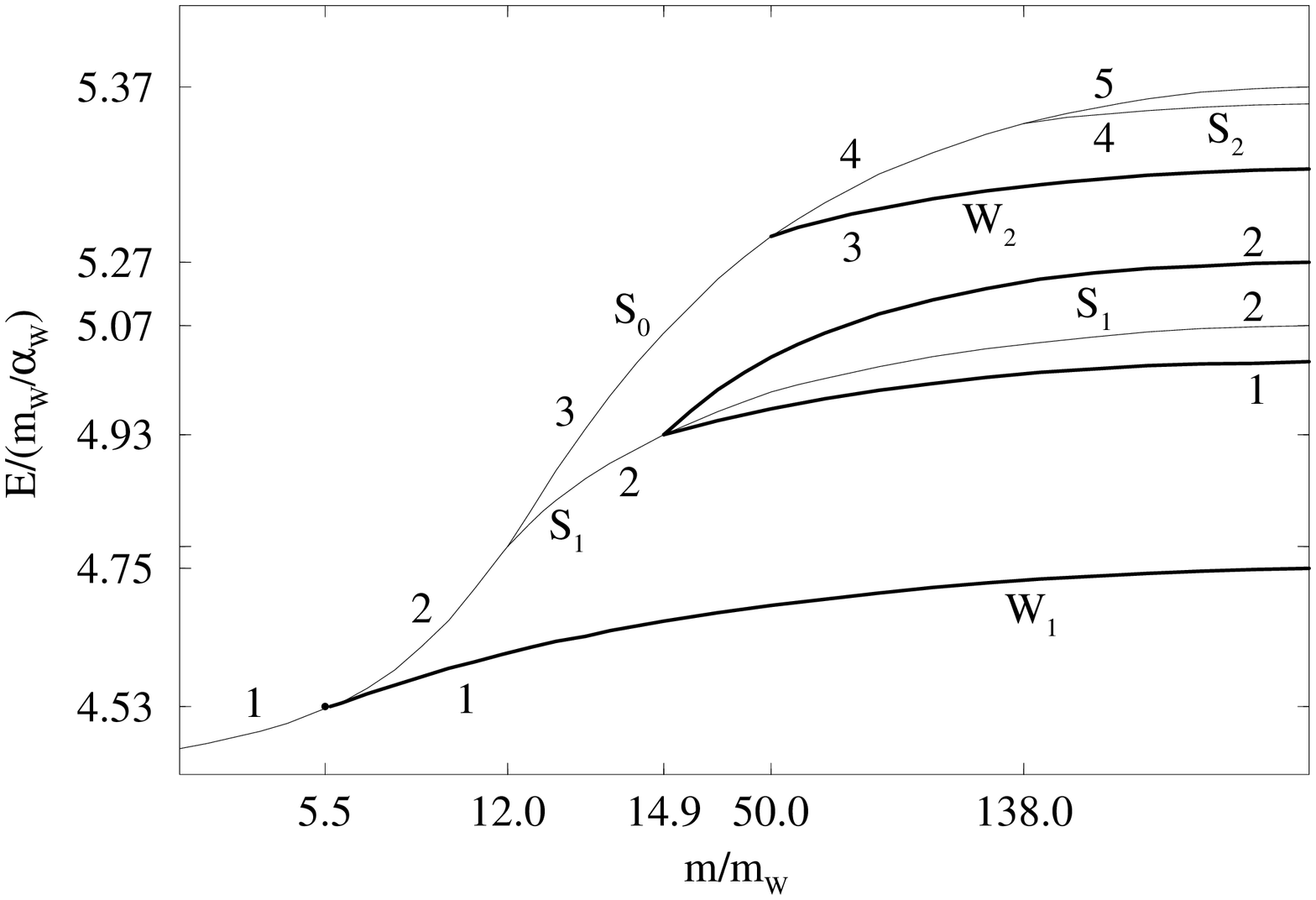,height=8cm,width=12.5cm,angle=270}
 \hskip 1.3cm
}}
\vspace {0.3cm}
{\bf Figure 2}. {\ninerm 
Solutions of the two-Higgs
model,  eq. (2),  for $ {\scriptstyle {\ninerm tan} \beta = 1}$, 
as function of the (common) radial Higgs mass $\scriptstyle m$.
Both the vertical and the horizontal axes have non-uniform scales. 
The number of negative modes is shown  explicitly for
 every  branch.
The winding branch,} ${\scriptstyle W_1}$, 
{\ninerm emerges out of the main sphaleron at }
${\scriptstyle m/m_W \simeq 5.5}$,
{\ninerm but remains unstable at least up to a
 mass ratio of $\scriptstyle \sim  200$.}

\vskip 0.5cm

\noindent  
with $n_s$ the number of unstable modes of the solution, is conserved
at all bifurcation points.  This is also true at $m/m_W \simeq 14.9$,
where a once- and a twice-unstable branch emerge out of the first
deformed sphaleron, corresponding to the transformation of a local
maximum, into two maxima separated by a lower-energy saddle.

The most interesting feature of Figure 2 is of course the winding
branch $W_1$, which on the basis of the 2d analog could have evolved
into a stable soliton at sufficiently large Higgs masses.  The
negative eigenvalue does not, however, cross the zero axis in this
case, at least all the way out to $m/m_W \sim 200$ (see table 1)
\footnote{
Such large values of parameters are in any case academic, since the
semiclassical approximation cannot be anymore trusted.}.  
By performing a numerical scan, we have checked that the winding
branch does not stabilize anywhere inside the larger space of
parameters of the potential, eq. (2) (see Table 2). Since we do not
expect interactions between the two doublets to improve stability, we
believe that such winding solitons do not exist. This is the main
conclusion of the present letter.

The unstable solutions are nevertheless of interest in their own
right. They emerge inside the region of weak coupling, and can be
lower in energy and comparatively less unstable than the known
sphaleron and deformed sphaleron solutions, as shown in Table 1.  They
furthermore continue to exist in a large region of parameter space,
which is sampled selectively in table 2.  The profile of a typical
solution for $\mbox{tan} \beta = 1.4$, $m_1 = 0.1$ and $m_2 = 10.1$ is
plotted in Fig. 3.  As is the case with deformed sphalerons in the
minimal model, the appearance of the two conjugate $W_1$ branches
corresponds to the splitting of a minimum-height passage in the energy
landscape into two.  These configurations should thus dominate
baryon-number violating transitions at high temperature, in the
appropriate regions of parameter space of the 2HSM.  More generally,
their existence illustrates how rich the configuration space of the
2HSM is. Its exploration could, we believe, reveal many surprising
features.

\vskip 0.5cm

\begin{center}
\begin{tabular}{|c|c|c||c|c||c|c|} \hline
          \hspace{0.cm} $ m_1=m_2 $ \hspace{0.cm} &
                \hspace{0.2cm} $ E_{W_1} $ \hspace{0.2cm} &
               \hspace{0.cm} $-\omega_{W_1}^2 $ \hspace{0.cm} &
               \hspace{0.2cm} $ E_{sph} $ \hspace{0.2cm} &
               \hspace{0.cm} $ -\omega_{sph}^2 $ \hspace{0.cm} &
               \hspace{0.1cm} $ E_{dsph} $ \hspace{0.1cm} &
               \hspace{0.cm} $ -\omega_{dsph}^2 $ \hspace{0.cm}
               \\ \hline
       6.0  &  4.53 &  5.17 & 4.53 & 6.35  &  --  &  --  \\ \hline
      10.0  &  4.67 &  3.09 & 4.78 & 11.21 &  --  &  --  \\ \hline
      15.0  &  4.72 &  2.68 & 4.95 & 23.43 & 4.93 & 8.44 \\ \hline 
      20.0  &  4.73 &  2.55 & 5.04 & 42.80 & 4.99 & 6.09 \\ \hline
      30.0  &  4.74 &  2.46 & 5.15 & 101.1 & 5.04 & 5.20 \\ \hline
      50.0  &  4.75 &  2.42 & 5.25 & 291.8 & 5.06 & 4.97 \\ \hline
     100.0  &  4.75 &  2.41 & 5.33 & 1192  & 5.07 & 4.75 \\ \hline
\end{tabular}
\end{center}

{\bf Table 1}. {\ninerm 
The energy (in units of ${\scriptstyle m_W/\alpha_W}$)
  and negative eigenvalue  of the fluctuation spectrum
 (in units of 
${\scriptstyle m_W}$)
along the winding  branch ${\scriptstyle W_1}$, in
the symmetric case
${\scriptstyle {\ninerm tan}\beta = 1}$ and ${\scriptstyle m_1=m_2}$.
We have included for comparison the energy and most negative
eigenvalue along  the embedded sphaleron  and first deformed
sphaleron branches, respectively ${\scriptstyle S_0}$ and
${\scriptstyle S_1}$.} 

\begin{center}
\begin{tabular}{|c|c|c||c|c|} \hline
               \hspace{0.3cm} $ m_1 $ \hspace{0.3cm} &
               \hspace{0.3cm} $ m_2 $ \hspace{0.3cm} &
               \hspace{0.3cm} $ tan\beta $ \hspace{0.3cm} &
               \hspace{0.5cm} $ E_{W_1}  $ \hspace{0.5cm} &
               \hspace{0.5cm} $ -\omega_{W_1}^2 $ \hspace{0.5cm} 
               \\ \hline\hline
       10.0  &  10.0  &   0.03  &  4.78   &  11.20  \\ \hline
       10.0  &  10.0  &   0.1  &  4.78   &  11.14  \\ \hline
       10.0  &  10.0  &   0.5  &  4.74   &   5.43  \\ \hline
       10.0  &  10.0  &   1.0  &  4.67   &   3.09  \\ \hline
       10.0  &  10.0  &   1.4  &  4.70   &   3.70  \\ \hline
       10.0  &  10.0  &   1.8  &  4.73   &   4.82  \\ \hline
       10.0  &  10.0  &   2.8  &  4.76   &   7.72  \\ \hline
       10.0  &  10.0  &   4.0  &  4.77   &   9.83  \\ \hline
       10.0  &  10.0  &   6.0  &  4.78   &  10.86  \\ \hline\hline
        6.0  &  6.0   &   2.0  &  4.53   &  5.95   \\ \hline
        5.0  &  6.0   &   2.0  &  4.51   &  6.05   \\ \hline
        4.0  &  6.0   &   2.0  &  4.49   &  5.94   \\ \hline
        3.0  &  6.0   &   2.0  &  4.46   &  5.85   \\ \hline
        2.5  &  6.0   &   2.0  &  4.44   &  5.80   \\ \hline
        2.0  &  6.0   &   2.0  &  4.43   &  5.76   \\ \hline
        1.5  &  6.0   &   2.0  &  4.40   &  5.72   \\ \hline\hline
       10.0  &  0.0   &   4.0  &  3.17   &  1.42   \\ \hline
       10.0  &  0.1   &   4.0  &  3.26   &  1.58   \\ \hline
       10.0  &  0.4   &   4.0  &  3.45   &  1.92   \\ \hline
       10.0  &  1.0   &   4.0  &  3.71   &  2.45   \\ \hline
       10.0  &  2.0   &   4.0  &  4.00   &  3.21   \\ \hline
       10.0  &  5.0   &   4.0  &  4.45   &  5.29   \\ \hline
       10.0  & 10.0   &   4.0  &  4.77   &  9.83   \\ \hline
       10.0  & 15.0   &   4.0  &  4.90   &  6.43   \\ \hline
       10.0  & 20.0   &   4.0  &  4.95   &  5.25   \\ \hline
       10.0  & 30.0   &   4.0  &  4.99   &  4.67   \\ \hline
       10.0  & 40.0   &   4.0  &  4.99   &  4.51   \\ \hline
\end{tabular}
\end{center}

{\bf Table 2}. {\ninerm 
The energy and negative eigenvalue of  the ${\scriptstyle W_1}$
 branch, for various values of  
${\scriptstyle {\ninerm tan}\beta}$ and of the ratio 
${\scriptstyle m_2/m_1}$. The units
are the same as in table 1. 
Notice that when  
${\scriptstyle {\ninerm tan}\beta \rightarrow 0}$ 
the winding solution merges with
the sphaleron.
For equal values of the two  Higgs masses the
lowest-energy,   least unstable solution is obtained in the symmetric
case  ${\scriptstyle {\ninerm tan}\beta = 1}$, when the winding is
equally shared by the two Higgses.
} 

\vspace {0.5cm}
\centerline{\hbox{ 
 \psfig{file=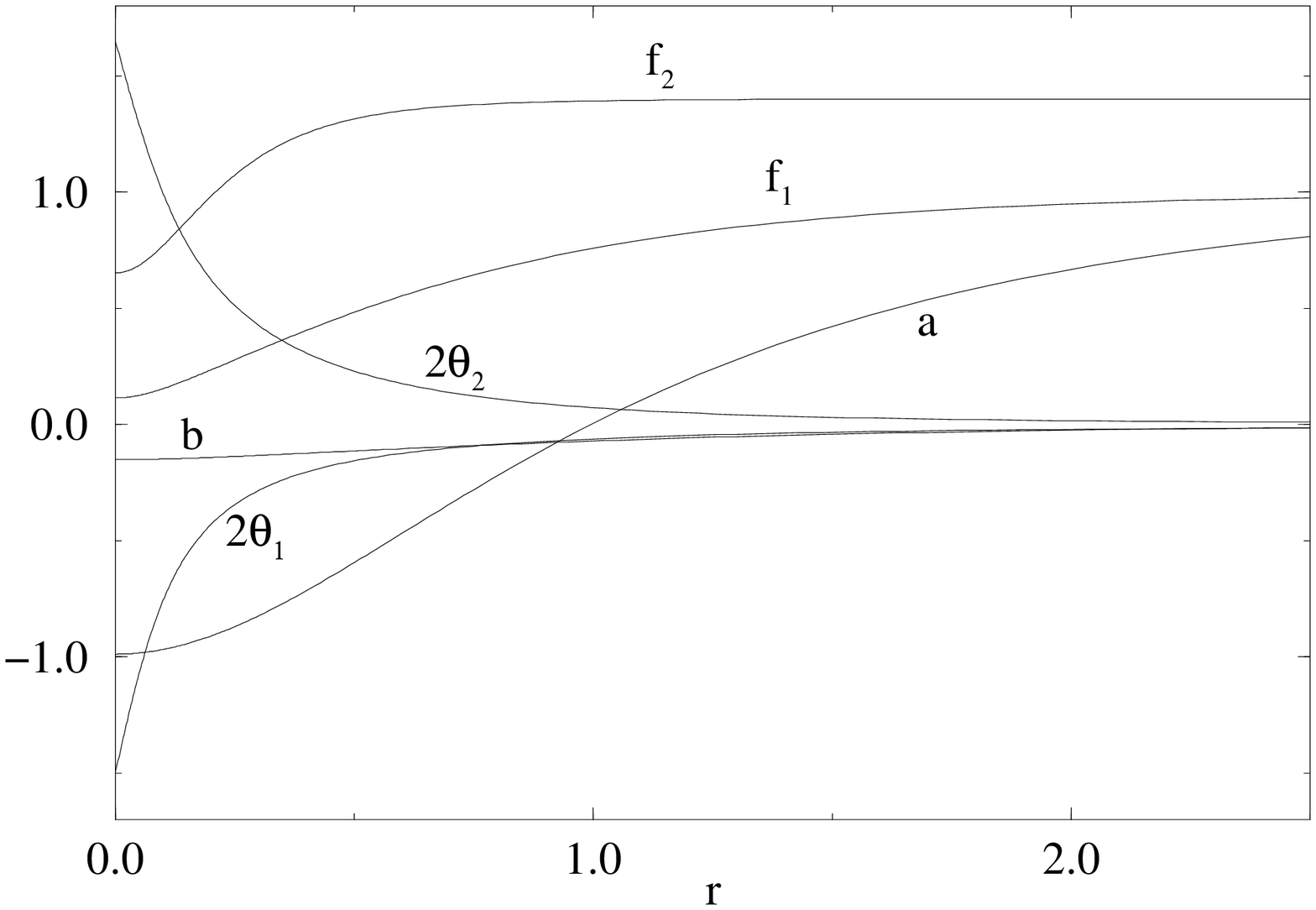,height=8cm,width=12.5cm,angle=270}
 \hskip 0.6cm
}}

\vspace {0.5cm}
{\bf Figure 3}. {\ninerm 
The profile of a solution on the ${\scriptstyle W_1}$ branch
for Higgs masses about 8 and 800 GeV respectively.}

\vspace{0.3cm}

\vskip 0.8cm

%Coomments:\hfil\break
%1)Ratio of masses? (kanena yeniko statement)\hfil\break
%2) Ehoume kanena data me $\lambda_3 \not= 0$? Inai i moni
%interaction pou boro na fadasto na  veltioni tin stability
%ton solitons \hfil\break
%3) i negative eigenvalue ston pinaka 4 afxani otan i maza
%m1 afxani. Inai numerically significant afto, ke an nai pos to
%katalavenis?\hfil\break

\vskip 1cm

\noindent
{\bf Acknowledgments} \\ We acknowledge a useful conversation with
Y. Brihaye.  C.B. and P.T. thank the Physics Department of the
University of Crete and the Research Center of Crete, while P.T. and
T.N.T. thank the Centre de Physique Th$\acute {\rm e}$orique of the
Ecole Polytechnique for hospitality, while this work was being
completed.  The work of C.B and TN.T.  was supported in part by the
EEC grants CHRX-CT93-0340 and CHRX-CT94-0621, and by the Greek General
Secretariat of Research and Technology grant No. 91E$\Delta$358, while
that of P.T.  by the INTAS grant INTAS-94-2352 and by the Russian
Foundatin for Fundamental Research, grant 96-02-17804a.

\vfill


\noindent

\vskip 0.5cm

\begin{thebibliography} {6666}

\bibitem{BT1} C. Bachas and T.N. Tomaras, Nucl. Phys. {\bf B428} 
(1994) 209
and Phys. Rev. {\bf D51} (1995) R5356.

\bibitem{BT3} C. Bachas and T.N. Tomaras, Phys. Rev. Lett. {\bf 76} 
(1996) 
356.

\bibitem{Turok} V. Kuzmin, V. Rubakov and M. Shaposhnikov, Phys. Lett.
{\bf 155B} (1985) 36; for recent  reviews  see
N. Turok, in {\it Perspectives on Higgs Physics}, G. Kane ed.,
World Scientific 1993,  and
V.A. Rubakov and M.E. Shaposhnikov, {\it Electroweak baryon number
non-conservation in the early Universe and in high energy collisions},
CERN-TH/96-13, hep-ph/9603208 (to appear in Usp. Fiz. Nauk, v.166, 
 May 1996).

\bibitem{Peccei} B. Kastening, R.D. Peccei and X. Zhang, 
 Phys. Lett. {\bf 266B} (1991) 413.

\bibitem{Bri} Y. Brihaye, J. Kunz and C. Semay, Phys. Rev. {\bf D42}
(1990) 193 and {\bf D44} (1991) 250.

\bibitem{spha} R. Dashen, B. Hasslacher and A. Neveu, 
 Phys. Rev. {\bf D10} (1974) 4138;
N.S. Manton, Phys. Rev. {\bf D28} (1983) 2019;\\
F. Klinkhamer and N.S. Manton, Phys. Rev. {\bf D30} (1984) 2212.

\bibitem{defspha}
J. Kunz and Y. Brihaye, Phys. Lett. {\bf 216B} (1989) 353.

\bibitem{Yaffe}
L. Yaffe, Phys. Rev. {\bf D40} (1989) 3463.

\bibitem{2d} N.S. Manton and T.M. Samols, Phys. Lett. {\bf 207B}
(1988) 179;\\
J.-Q. Liang, H.J.W. M\"uller-Kirsten and D.H. Tchrakian, Phys. Lett.
{\bf 282B} (1992) 105.

\bibitem{2dab} Y. Brihaye, S. Giller, P. Kosinski and J. Kunz,
 Phys. Lett. {\bf 293B} (1992) 383.

\bibitem{Numerics} 
W. Press, S. Teukolsky, W. Vetterling and B. Flannery, {\it Numerical
Recipes}, chapter 17. Cambridge University Press 1992.

\end{thebibliography}
\end{document}